\PassOptionsToPackage{unicode=true}{hyperref} 
\PassOptionsToPackage{hyphens}{url}
\documentclass[
  format=sigconf,
  review=false,
  timestamp=true,
  authordraft=false,
  letterpaper,
  nonacm=true]{acmart}
\usepackage[]{amsmath}
\usepackage{amssymb,amsmath}
\usepackage{ifxetex,ifluatex}
\ifnum 0\ifxetex 1\fi\ifluatex 1\fi=0 
  \usepackage[T1]{fontenc}
  \usepackage[utf8]{inputenc}
  \usepackage{textcomp} 
\else 
  \usepackage{unicode-math}
  \defaultfontfeatures{Scale=MatchLowercase}
  \defaultfontfeatures[\rmfamily]{Ligatures=TeX,Scale=1}
\fi
\IfFileExists{upquote.sty}{\usepackage{upquote}}{}
\IfFileExists{microtype.sty}{
  \usepackage[]{microtype}
  \UseMicrotypeSet[protrusion]{basicmath} 
}{}
\makeatletter
\@ifundefined{KOMAClassName}{
  \IfFileExists{parskip.sty}{%
    \usepackage{parskip}
  }{
    \setlength{\parindent}{0pt}
    \setlength{\parskip}{6pt plus 2pt minus 1pt}}
}{
  \KOMAoptions{parskip=half}}
\makeatother
\usepackage{xcolor}
\IfFileExists{xurl.sty}{\usepackage{xurl}}{} 
\IfFileExists{bookmark.sty}{\usepackage{bookmark}}{\usepackage{hyperref}}
\hypersetup{
  pdftitle={Infrastructure-Agnostic Hypertext},
  pdfkeywords={Xanadu, hypermedia, transclusion, documents},
  pdfborder={0 0 0},
  breaklinks=true}
\providecommand{\tightlist}{%
  \setlength{\itemsep}{0pt}\setlength{\parskip}{0pt}}
\ifx\paragraph\undefined\else
  \let\oldparagraph\paragraph
  \renewcommand{\paragraph}[1]{\oldparagraph{#1}\mbox{}}
\fi
\ifx\subparagraph\undefined\else
  \let\oldsubparagraph\subparagraph
  \renewcommand{\subparagraph}[1]{\oldsubparagraph{#1}\mbox{}}
\fi

\makeatletter
\def\fps@figure{htbp}
\makeatother

\usepackage{tikz}
\usetikzlibrary{shapes}
\usetikzlibrary{positioning}
\usepackage[]{natbib}
\title{Infrastructure-Agnostic Hypertext}
\date{2019-06-28}

\begin{document}

\begin{abstract}
This paper presents a novel and formal interpretation of the original
vision of hypertext: infrastructure-agnostic hypertext is independent
from specific standards such as data formats and network protocols. Its
model is illustrated with examples and references to existing
technologies that allow for implementation and integration in current
information infrastructures such as the Internet.
\end{abstract}

\author{Jakob Voß}
\email{jakob.voss@gbv.de}
\orcid{0000-0002-7613-4123}
\affiliation{%
\institution{Verbundzentrale des GBV (VZG)}
}


\maketitle

\hypertarget{introduction}{%
\section{Introduction}\label{introduction}}

The original vision of hypertext as proposed by Ted Nelson
\citep{Nelson1965, Nelson2007} still waits to be realized. His influence
is visible through people who, influenced by his works, shaped the
computer world of today, last but not least the Web
\citep{Nelson2008}.\footnote{Tim Berners-Lee references Nelson both in
  the proposal that led to the Web \citep{BernersLee1990, Nelson1967}
  and at the early W3C homepage \url{http://www.w3.org/Xanadu.html}
  (1992/93) \citep{Nelson1980}.} Nelson's core idea, a network of
visibly connected documents called Xanadu, goes beyond the Web in
several aspects. In particular it promises non-breaking links and it
uses links to build documents (with versions, quotations, overlay
markup\ldots{}) instead of using documents to build links
\citep{Nelson1997}. The concept of hypertext or more general
`hypermedia' has also been used differently from Nelson, both in the
literary community (that focused on simple links), and in the hypertext
research community (that focused on tools) \citep{WardripFruin2004}.

This paper tries to get back to the original vision of hypertext by
specification of a formal model that puts transcludeable documents at
its heart. Apart from Nelson's works
\citep{Nelson1965, Nelson1967, Nelson1974, Nelson1980, Nelson1999, Nelson2007}
this paper draws from research on data format analysis
\citep{Voss2013a, Kent1989}, content-based identifiers
\citep{Trask2016, Lukka2002}, existing transclusion technologies for the
Web \citep{Akscyn2015, Tennison2011, IIIFImageAPI, Csillag2013} and
hypertext systems beyond link-based models \citep{Atzenbeck2017}.

\begin{figure}
\Description{Several data formats connected by arrows}
\scalebox{0.9}{\usetikzlibrary{positioning}
\begin{tikzpicture}[font=\sffamily]

\node (markdown) {Markdown};
\node [right=15mm of markdown] (tex) {\TeX};
\node [right=15mm of tex] (pdf) {\textbf{PDF}};

\node [below=6mm of markdown] (tikz) {Ti\textit{k}Z};
\node [right=of tikz] (svg) {SVG};
\node [below=6mm of pdf] (html) {\textbf{HTML}};

\node [below=13mm of tikz] (wikidata) {Wikidata};
\node [right=of wikidata] (csljson) {CSL JSON};
\node [right=of csljson] (bibtex) {Bib\negthinspace\TeX};

\node [above=4mm of tex] (textemplate) {Pandoc template (\TeX)};
\node [below=4mm of svg,xshift=-6mm] (htmltemplate) {Pandoc template (HTML)};
\node [below=5mm of html] (css) {CSS};

\draw[->] (markdown) to (tex);
\draw[->] (markdown) to (html);
\draw[->] (tex) to (pdf);
\draw[->] (tikz) to (svg);
\draw[dotted,->] (svg) to (html);
\draw[dotted,->] (css) to (html);
\draw[dotted,->] (tikz) to (tex);
\draw[->] (wikidata) to (csljson);
\draw[->] (textemplate) to (tex);
\draw[->] (htmltemplate) to (html);
\draw[->] (csljson) to (bibtex);
\draw[->,bend angle=90, bend right] (bibtex) to (pdf);
\draw[->] (csljson) to (html);

\end{tikzpicture}}
\caption{Proto-transclusion links of this paper}
\label{fig:demopaper}
\end{figure}
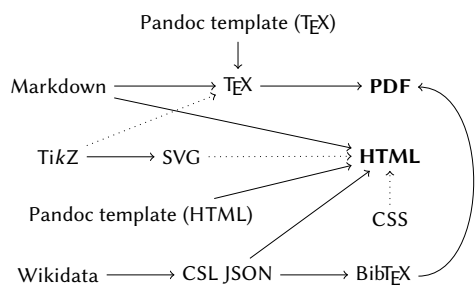

Limited by the state of document processing tools and by submission
guidelines, this paper is not a demo of hypertext.\footnote{See
  \citep{Capadisli2015} for a good example of what an actual demo paper
  might look like.} On a closer look however there are traces of
transclusion links that have been processed to this paper. See figure
\ref{fig:demopaper} for an overview and
\url{https://github.com/jakobib/hypertext2019} for details and sources.

\hypertarget{outline}{%
\section{Outline}\label{outline}}

The architecture of an infrastructure-agnostic hypertext system consists
of four basic elements and their relations:

\begin{enumerate}
\def\labelenumi{\arabic{enumi})}
\item
  \textbf{documents} include all finite, digital objects
\item
  \textbf{document identifiers} reference individual documents
\item
  \textbf{content locators} reference segments within documents
\item
  \textbf{edit list} combine parts of existing document into new ones
\end{enumerate}

Instances of each element can further be grouped by \textbf{data
formats}. The elements and their relations are each described in the
following sections after a formal definition.

\hypertarget{formal-model}{%
\subsection{Formal model}\label{formal-model}}

A hypertext system is a tuple \(\langle D,I,C,E,S,R,U,T,A \rangle\)
where:

\begin{description}
\tightlist
\item[\(D\)]
is a set of documents
\item[\(I\)]
is a set of document identifiers with \(I \subset D\)
\item[\(C\)]
is a set of content locators with \(C \subset D\)
\item[\(E\)]
is a set of edit lists with \(E \subset D\)
\item[\(S\)]
is a set of document segments with \(S \subset C \times D\)
\end{description}

Document sets can each be grouped into (possibly overlapping) data
formats. The hypertext system further consists of:

\begin{description}
\tightlist
\item[\(R\)]
is a retrieval function with \(R\colon I \to D\)
\item[\(U\)]
is a segments usage function \(U\colon E \to \mathcal{P}(S)\)
\item[\(T\)]
is a transclusion function with \(T\colon S \to D\)
\item[\(A\)]
is an hypertext assemble function with \(A\colon E \to D\)
\end{description}

A practical hypertext system needs executable implementations of the
functions \(R,U,T,A\) and a method to tell whether a given combination
\(\langle c,d \rangle \in C \times D\) is part of \(S\) to allow its use
with \(T\).

\hypertarget{documents}{%
\subsection{Documents}\label{documents}}

A document is a finite sequence of bytes. This definition roughly
equates with the definition of data as documents
\citep{Furner2016, Voss2013b}. The notion of hypertext used in this
paper therefore subsumes all kinds of hyperdata (datasets that
transclude other datasets). Documents are static by content
\citep{Renear2009} but they may be processed dynamically. Documents are
grouped by non-disjoint data formats such as UTF-8, CSV, SVG, PDF and
many many more.

\hypertarget{document-identifiers}{%
\subsection{Document identifiers}\label{document-identifiers}}

A document identifier is a relatively short document that refers to
another document. Identifiers have properties depending on the
particular identifier system they belong to
\citep[pp.~59-71]{Voss2013b}. Identifiers in infrastructure-agnostic
hypertext must first be \emph{unambiguous} (an identifier must reference
only one document), \emph{persistent} (the reference must not change
over time), and \emph{actionable} (hypertext systems should provide
methods to retrieve documents via the retrieval function \(R\)).
Properties that should be fulfilled at least to some degree include
\emph{uniqueness} (a document should not be referenced by too many
identifiers), \emph{performance} (identifiers should be easy to compute
and to validate), and \emph{distributedness} (identifiers should not
require a central institution). The actual choice of an identifier
system depends on weighting of specific requirements. A promising choice
is the application of content-based identifiers as proposed at least
once in the context of hypertext systems \citep{Lukka2002}.

\hypertarget{content-locators}{%
\subsection{Content locators}\label{content-locators}}

A content locator is a document that can be used to select parts of
another document via transclusion. Nelsons refers to these locators as
``reference pointers'' \citep{Nelson1999}, exemplified with spans of
bytes or characters in a document. Content locators depend on data
formats and document models. For instance locator languages XPath,
XPointer, and XQuery act on XML documents, which can be serialized in
different forms (therefore it makes no sense to locate parts of an XML
document with positions of bytes). Other locator languages apply to
tabular data (SQL, RFC~7111), to graphs (SPARQL, GraphQL), or to
two-dimensional images (IIIF), to name a few. Whether and how parts of a
document can be selected with a content locator language depends on
which data format the document is interpreted in. For instance an SVG
image file can be processed at least as image, as XML document, or as
Unicode string, each with its own methods of locating document segments.
Content locators can be extended to all executable programs that
reproducibly process some documents into other documents. This
generalization can be useful to track data processing pipelines as
hyperdata such as discussed for executable papers and reproducible
research. Restriction of content locators to less powerful query
languages might make sense from a security point of view.

\hypertarget{edit-lists}{%
\subsection{Edit Lists}\label{edit-lists}}

An edit list is a document that describes how to construct a new
document from parts of other documents. Edit lists are known as
\emph{Edit Decision Lists} in Xanadu and the idea was borrowed from film
making \citep{Nelson1967}. Simplified forms of edit lists are
implemented in version control systems and in collaborative tools such
as wikis and real-time editing. Hypertext edit lists go beyond this
one-dimensional case by support of multiple source documents and by more
flexible methods of document processing in addition to basic operations
such as insert, delete, and replace. The actual processing steps tracked
by an edit list depend on data formats of transcluded documents. Just
like content locators, edit lists could be extended to arbitrary
executable programs that implement the hypertext assemble function \(A\)
for some subset of edit lists. To ensure reproducibility and reliable
transclusion,\footnote{If documents models/segments change, locators may
  not be applicable anymore \citep{Csillag2013}.} these programs must
not access unstable external information such as documents that may
change \citep{Renear2009}.

\hypertarget{data-formats}{%
\subsection{Data formats}\label{data-formats}}

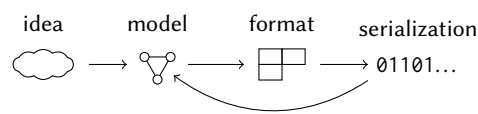
\begin{figure}
\Description{Three icons labeled `ideas`, `model`, `format`, and `serialization`, respectively}
\usetikzlibrary{shapes}
\usetikzlibrary{positioning}
\begin{tikzpicture}[font=\sffamily,node distance=4mm,label distance=1.25mm]
\matrix[column sep=6mm] {
  \node[draw,cloud,cloud puffs=10,cloud ignores aspect,cloud puff arc=90,
        minimum width=8mm,minimum height=4mm,label=idea] (idea) {};
&
 \node[ellipse,minimum width=6mm,label=model,minimum height=4mm] (model) {};
 \draw (30:2mm) -- (-90:2mm) -- (150:2mm) -- cycle;
 \draw [draw,fill=white]  (-90:2mm) circle (.6mm);
 \draw [draw,fill=white] (30:2mm) circle (.6mm);
 \draw [draw,fill=white] (150:2mm) circle (.6mm);
&
 \node[rectangle,minimum width=6mm,minimum height=4mm,label=format] (schema) {};
 \draw (schema.north west) rectangle (schema.center);
 \draw (schema.north) rectangle (schema.east);
 \draw (schema.west) rectangle (schema.south);
&
 \node[rectangle,minimum height=4mm] (data) {\texttt{01101}\ldots}; \\
};
\node[above=0.5mm of data] {serialization};
\draw[->,shorten >=1mm,shorten <=2mm] (idea) to (model);
\draw[->,shorten >=2mm,shorten <=1mm] (model) to (schema);
\draw[->,shorten >=0mm,shorten <=2mm] (schema) to (data);
\draw[<-,bend angle=35, bend right] (model.south east) to (data.south west);
\end{tikzpicture}
\caption{levels of data modeling}
\label{fig:datamodeling}
\end{figure}

An often neglected fundamental property of digital documents is their
grounding in data formats. A data format is a set of documents that
share a common data model, also known as their document model, and a
common serialization (see fig.~\ref{fig:datamodeling}). Models define
elements of a document in terms of sets, strings, tuples, graphs or
similar structures. These structures are mathematically rigorous in
theory \citep{Renear2009} but more based on descriptive patterns in
practice \citep{Voss2013a}. The meaning of these elements (for instance
``words'', ``sentences'', and ``paragraphs'' in a document model) is
based on ideas that we at least assume to be consistent among different
people.

Data modeling, the act of mapping between ideas, models, and formats is
an unsolved problem because ideas can be expressed in many models and
models can be interpreted in many ways \citep{Kent1989}. Data models can
further be expressed in multiple formats although these formats should
fully be convertible between each other, at least in theory.\footnote{In
  practice it's often unknown whether two data formats actually share
  the same model, especially if models are only given implicitly by
  definition of their formats.} Formats can further be serialized in
multiple forms, which for their part are based on other data models. For
instance the RDF model can be serialized in RDF/XML format which is
based on the XML model. XML can be serialized XML syntax which is based
on the Unicode model, and Unicode can be serialized in UTF-8. At the end
of these chains of abstraction eventually all documents can be
serialized as sequence of bytes. Serializations are seldom simple
mappings as most serialization formats allow insignificant variances
such as additional whitespace. To check whether a data object conforms
to a serialization, formats are often described with a formal grammar
that can also give insights about the format's model.\footnote{Formats
  may also exist purely implicit in form of sample instances which
  grammar, model, and ideas must be guessed from by reverse engineering
  \citep{Voss2013a}.} Infrastructure-agnostic hypertext does not impose
any limits on possible data formats and their models.

\hypertarget{example}{%
\section{Example}\label{example}}

The following example may illustrate the formal model and its elements.
Let \(\langle D,I,C,E,S,R,U,T,A \rangle\) be a hypertext system with
\(D\) the set of printable ASCII character strings and some documents:

\begin{description}
\tightlist
\item[\(d_1\)]
= `\texttt{My\ name\ is\ Alice}'
\item[\(d_2\)]
= `\texttt{Alice}'
\item[\(c_1\)]
= `\texttt{char=11,15}'
\item[\(d_3\)]
= `\texttt{Hello,\ !}'
\item[\(c_2\)]
= `\texttt{char=7}'
\item[\(d_4\)]
= `\texttt{Hello,\ Alice!}'
\end{description}

If \(c_1\) and \(c_2\) are read in content locator syntax as defined by
RFC~5147 so that \(T(\langle c_1,d_1 \rangle) = d_2\) then \(d_4\) can
be constructed by transcluding a document segment of \(d_1\) into
\(d_3\) at position \(c_2\). The corresponding edit list \(e_1 \in E\)
with \(A(e_1)=d_4\) could look like this:

\begin{verbatim}
 take      995f37f2e066b7d8893873ca4d780da5bf017184
 insert at 48ba94c47b45390b6dd27824cfc0d8468c2cbc71
 from      fcb59267e2e6641140578235c8cb6d38eaf6abc1
 segment   c5b794c7ae5d490f52a414d9d19311b9a19f61b3
\end{verbatim}

The values in \(e_1\) are SHA-1 hashes of \(d_3\), \(c_2\), \(d_1\), and
\(c_1\) respectively.\footnote{A more practical edit list syntax \(E\)
  could also allow the direct embedding of small document instances
  which SHA-1 hashes can be computed from. If implemented carefully,
  this could also reconcile transclusion with copy-and-paste.} Retrieval
function \(R\) maps them back to strings. Hyperlinks are given by
\(U(e_1) = \{ \langle c_2,d_3 \rangle, \langle c_1,d_1 \rangle \}\) used
for editing \(d_3\) to \(d_4\) (versioning) and for referencing of
segment of \(d_1\) in \(d_4\) (transclusion).

\hypertarget{implementations}{%
\section{Implementations}\label{implementations}}

One of the problems faced by project Xanadu was it long required new
developments such as computer networks, document processing, and
graphical user interfaces ahead of their time. Today we can build on a
many existing technologies:

\begin{description}
\tightlist
\item[networks:]
storage and communication networks are ubiquitous with several protocols
(HTTP, IPFS, BitTorrent\ldots{}).
\item[identifier systems:]
document identifiers should be part of the URI/IRI identifier system.
More specific candidates of relevant identifier systems include URLs and
content-based identifiers.
\item[formats:]
hypertext systems should not be limited to their own document formats
(such as the Web's focus on HTML/DOM) but allow for integration of all
kinds of digital objects.
\item[content locators]
as shown above, several content locator and query formats exist, at
least for some document models.
\end{description}

Access to documents via a retrieval function \(R\) can be implemented
with existing network and identifier technologies. Obvious solutions
build on top of HTTP and URL but these identifiers are far from
unambiguous and persistent. Content-based identifiers are guaranteed to
always reference the same document but they require network and storage
systems to be actionable.\footnote{New standards such as IPFS Mulihashes
  and BitTorrent Merkle-Hashes look promising but these types of
  identifiers are not specified as part of the URI system (yet)
  \citep{Trask2016}.} The set of supported data formats is only limited
by availability of applications to view and to edit documents. Full
integration into a hypertext system however requires appropriate content
locator formats to select, transclude, and link to segments from these
documents. Existing content locator technologies include URI Fragment
Identifiers \citep{Tennison2011}, patch formats (JSON Patch, XML Patch,
LD Patch\ldots{}), and domain-specific query languages as long as they
can guarantee reproducible builds. The IIIF Image API, with focus on
content locators in images \citep{IIIFImageAPI}, and hypothes.is, with a
combination of locator methods \citep{Csillag2013}, popularized at least
simple forms of transclusion on the Web.

\hypertarget{challenges}{%
\subsection{Challenges}\label{challenges}}

Despite the availability of technologies to build on, creation of a
xanalogical hypertext system is challenging for several reasons. The
general problems involved with transclusion have been identified
\citep{Akscyn2015}. Other or more specific challenges include (ordered
by severity):

\begin{description}
\tightlist
\item[storage:]
data storage is cheap, but someone has to pay for it.
\item[normalization:]
most documents (including identifiers) can be serialized in different
forms. To support unique document identifiers, a hypertext system should
support normalization of documents to canonical forms.
\item[link services:]
databases of links have been proposed as central part of Open Hypermedia
Systems \citep{Atzenbeck2017} but they are not available for the Web
because of commercial interest.\footnote{In particular by search engines
  and by spammers. A criterion to judge the success of a hypertext
  system is whether it is popular enough to attract link spam.} Links
are ideally derived from edit lists with segments usage function \(U\).
Recent development such as Webmention and OpenCitation may help to
improve collection of links.
\item[visualization and navigation:]
this most recognizable element of hypertext has mostly been reduced to
simple links while Nelson's ideas seem to have been forgotten or ignored
\citep{Viegas2003}. Nevertheless the creation of tools for visualization
and navigation in hypertext structures is less challenging then getting
hold of the underlying documents and hyperlinks.
\item[edit list formats]
despite edit lists being the very core of the idea of hypertext
\citep{Nelson1967}, they have rarely been implemented in reusable data
formats. Proper hypertext implementations therefore require to establish
new formats with support of hypertext assemble function \(A\) and
segments usage function \(U\).
\item[editing tools]
applications to create and modify digital objects track changes don't
provide this information in form of reusable edit lists, if at all.
Hypermedia authoring needs to be integrated into existing editing tools
to succed \citep{DiIorio2005}.
\item[copyright and control:]
who should be allowed to use which documents under which conditions? The
answers primarily depend on legal, social, and political requirements.
\end{description}

\hypertarget{differences-to-xanadu}{%
\subsection{Differences to Xanadu}\label{differences-to-xanadu}}

Project Xanadu promised a comprehensive hypertext system including
elements for content (xanadocs), network (servers), rights
(micropayment), and interfaces (viewers) -- years before each these
concept made it into the computer mainstream. Today a xanalogical
hypertext system can more build on existing technologies. The
infrastructure-agnostic model of hypertext tries to capture the core
parts of the original vision of hypertext by concentrating on its
documents and document formats. For this reason some requirements listed
by Xanadu Australia \citep{Pam2002} or mentioned by Nelson in other
publications are not incorporated explicitly:

\begin{enumerate}
\def\labelenumi{\alph{enumi})}
\tightlist
\item
  identified servers as canonical sources of documents
\item
  identified users and access control
\item
  copyright and royalty system via micropayment
\item
  user interfaces to navigate and edit hypertexts
\end{enumerate}

Meeting these requirements in actual implementations is possible
nevertheless. Identified servers (a) and users (b) were part of
\emph{Tumbler} identifers (that combined document identifiers and
content locators) \citep{Nelson1980} but the current OpenXanadu
implementation uses plain URLs as part of its \emph{Xanadoc} edit list
format \citep{Nelson2014}. Canonical sources of documents (a) could also
be implemented by blockchains or alternative technology to prove that a
specific document existed on a specific server at a specific time. Such
knowledge of a document's first insertion into the hypertext system
would also allow for royalty systems (c).\footnote{Copyright detection
  was easy to implement with mandatory registration such as partly
  required in the United States until 1976. Authors might also register
  documents with cryptographic hashes without making them public in the
  first place.} Identification of users and access control (b) could
also be implemented in several ways but this feature much more depends
on network infrastructures and socio-technical environments, including
rules of privacy, intellectual property, and censorship. Last but not
least, a hypertext system needs applications to visualize, navigate, and
edit hypermedia (d)\footnote{Tim Berners-Lee's first Web browser
  originally supported editing.} Several user interface have been
invented in the history of hypertext \citep{MProve2002} and there will
unlikely be one final application because user interfaces depend on
use-cases and file formats.

\hypertarget{differences-to-other-hypertext-models}{%
\subsection{Differences to other hypertext
models}\label{differences-to-other-hypertext-models}}

The focus of models from the hypertext research community
\citep{Atzenbeck2017} is more on services and tools than on Nelson's
requirements \citep{WardripFruin2004}. This paper rather looks at the
neglected ``within-component layer''\footnote{``It would be folly to
  attempt a generic model covering all of these data types.''
  \citep{Halasz1990}} of the Dexter Hypertext Reference Model
\citep{Halasz1990} than on issues of storage, presentation and
interaction with a hypertext system. Extension with content locators
(``locSpecs'' in \citep{Gronbaeck1996}) could more align Dexter with
infrastructure agnostic hypertext but existing models rarely put
traceable edit-lists and transclusion into their core.

\hypertarget{summary-and-conclusion}{%
\section{Summary and conclusion}\label{summary-and-conclusion}}

This paper presents a novel interpretation of the original vision of
hypertext \citep{Nelson1965, Nelson1967}. Its infrastructure-agnostic
model does not require or exclude specific data formats or network
protocols. Abstract from these ever-changing technologies, the focus is
on hypermedia \emph{content} (documents) and \emph{connections}
(hyperlinks). Core elements of hypertext systems are identified as
documents, document identifiers, content locators, and edit lists. A
formal model defines their relations based on knowledge of data formats
and models. It is shown which technologies can be used to implement such
a hypertext system integrated into current information infrastructures
(especially the Internet and the Web) and which challenges still exist
(in particular support of edit lists in editing tools).

  \bibliography{infrastructure-agnostic-hypertext.bib}

\end{document}